\documentclass[twocolumn]{article}

\usepackage{graphicx}
\usepackage{dcolumn}
\usepackage{bm}
\usepackage{geometry,amsmath,amssymb} 
\newcommand{\tmmathbf}[1]{\ensuremath{\boldsymbol{#1}}}

\newcommand{\tmtextbf}[1]{{\bfseries{#1}}}
\usepackage[cm]{fullpage}
\usepackage{bbold}
\usepackage{sfmath}
\usepackage[font=small,sf,textfont=sf]{caption}
\usepackage[sf,bf,small,center]{titlesec}

\title{\textsf{Nonnormal amplification in random balanced neuronal networks}}
\author{Guillaume Hennequin \and Tim P. Vogels \and Wulfram Gerstner}

\date{\small School of Computer and Communication Sciences and Brain Mind Institute\\
Ecole Polytechnique F\'ed\'erale de Lausanne -- 1015 EPFL, Switzerland\\
Correspondence: \texttt{guillaume.hennequin@epfl.ch}\\
\vspace{1cm}
\hrule}

\begin{document}

\maketitle

\begin{abstract}
In dynamical models of cortical networks, the recurrent connectivity can
amplify the input given to the network in two distinct ways.  One is induced by
the presence of near-critical eigenvalues in the connectivity matrix
$\tmmathbf{W}$, producing large but slow activity fluctuations along the
corresponding eigenvectors (dynamical slowing). The other relies on
$\tmmathbf{W}$ being nonnormal, which allows the network activity to make large
but fast excursions along specific directions.  Here we investigate the
tradeoff between nonnormal amplification and dynamical slowing in the
spontaneous activity of large random neuronal networks composed of excitatory
and inhibitory neurons.  We use a Schur decomposition of $\tmmathbf{W}$ to
separate the two amplification mechanisms. Assuming linear stochastic dynamics,
we derive an exact expression for the expected amount of purely nonnormal
amplification.  We find that amplification is very limited if dynamical slowing
must be kept weak. We conclude that, to achieve strong transient amplification
with little slowing, the connectivity must be structured. We show that
unidirectional connections between neurons of the same type together with
reciprocal connections between neurons of different types, allow for
amplification already in the fast dynamical regime.  Finally, our results also
shed light on the differences between balanced networks in which inhibition
exactly cancels excitation, and those where inhibition dominates. 
\end{abstract}

\section{Introduction}                          

A puzzling feature of cortical dynamics is the presence of structure in
spontaneously generated activity states. For example, activity in cat primary
visual cortex fluctuates along some non-random spatial patterns even when
recordings are performed in complete darkness \cite{Tsodyks99,Kenet03}. 
Similarly, spontaneously generated patterns of firing rates in rat sensory cortices
occupy only part of the total space of theoretically possible patterns
\cite{Luczak09}. As the constraints that govern these dynamics cannot be
attributed to external stimuli, they are thought to originate from
the patterns of synaptic connectivity within the network \cite{Goldberg04,Murphy09}. 
This phenomenon is called patterned amplification.
                                               
Patterned amplification can also be observed in simulated neuronal networks,
in which spontaneous activity can be modelled as the response to unspecific,
noisy inputs delivered to each neuron individually.
Propagated through recurrent connections, these noisy inputs may cause the
activity of some neurons to transiently deviate from their average more strongly
than could be expected from the variability of the external inputs. 
We thus define amplification here as the strength of these additional,
connectivity-induced fluctuations.

Let us consider the following simple linear model for stochastic network dynamics:
\begin{equation}
  d \tmmathbf{x} = \frac{d t}{\tau} \left( \tmmathbf{W} - \mathbb{1}
  \right)  \tmmathbf{x} + \sigma_{\xi} d \tmmathbf{\xi} \label{ouprocess}
\end{equation}
where $\tau$ is the neuronal time constant, $\tmmathbf{x}\in \mathbb{R}^N$ is the deviation
of momentary network activity with respect to a constant mean firing rate,
$\tmmathbf{W}$ is an $N \times N$
synaptic connectivity matrix, $\mathbb{1}$ is the identity matrix, and
$d \tmmathbf{\xi}$ is a noise term modelled as a Wiener process. 
The fluctuations of $x_i(t)$ around zero (i.e. around the mean
firing rate of neuron $i$) is caused by the noisy input and the recurrent drive.
Starting from arbitrary initial conditions, the network activity $\tmmathbf{x}$ converges to
a stationary Gaussian process with covariance matrix $\tmmathbf{\Sigma} =
\left\{ \sigma_{i j} \right\}$ (at zero time lag), provided no eigenvalue of
$\tmmathbf{W}$ has a real part greater than unity.
This covariance matrix has a baseline component
$\Sigma_{\text{unc.}} = \sigma_\xi^2 \tau \mathbb{1}/2$ that corresponds
to the covariance matrix in the absence of network connections
(``unconnected'').  Wiring up the network yields additional correlations and
gives potentially rise
to larger fluctuations of the activity of individual units. We define this
amplification $A$ as the ratio $\left[\text{Tr}(\tmmathbf{\Sigma}) -
\text{Tr}(\tmmathbf{\Sigma}_{\text{unc.}}) \right] /
\text{Tr}(\tmmathbf{\Sigma}_{\text{unc.}})$.  In other words, $A$ measures the
relative gain in mean variance that can be attributed to the recurrent
connections.  That is, 
\begin{equation}
  A \left( \tmmathbf{W} \right) \hspace{1em} \overset{\text{def}}{=}
  \hspace{1em} \left[\frac{2}{\tau \sigma_{\xi}^2 N} \sum_{i = 1}^N \sigma_{i i}\right] -1
  \label{amplificationdef}
\end{equation}

Under linear dynamics like that of Eq.\ \ref{ouprocess}, amplification can
originate from two separate mechanisms.  A first, ``normal'' type of
amplification can arise from eigenvalues of
$\tmmathbf{W}$ with real parts close to (but smaller than) 1.
The noise accumulates along the associated eigenvectors more than
in other directions, giving rise to larger activity fluctuations
and substantial dynamical slowing along those axes.
If the synaptic connectivity is normal in the mathematical sense ($\tmmathbf{W} 
\tmmathbf{W}^{\dag} = \tmmathbf{W}^{\dag} \tmmathbf{W}$), it is the
\emph{only} mechanism through which the network can amplify its input \cite{Murphy09}.
Indeed, if $\tmmathbf{W}$ is normal, its eigenvectors form an orthonormal basis.
The sum of variances in this eigenbasis is therefore equal to the sum of variances of the
neuronal activities in the original equations. Since linear stability imposes that every eigenvalue of
$\tmmathbf{W}$ has a real part less than one, the activity along the eigenvectors
can only decay following some initial perturbation. In other words, a stable
\emph{normal} linear system is contractive: no initial condition can
transiently be amplified. If the matrix $\tmmathbf{W}$ is not normal
($\tmmathbf{W}  \tmmathbf{W}^{\dag} \neq \tmmathbf{W}^{\dag} \tmmathbf{W}$),
another, \emph{nonnormal} type of amplification can also
contribute \cite{Murphy09,Ganguli08,Goldman09,Trefethen05}. The eigenvectors
are no longer orthogonal to each other, and the apparent decay of the activity
in the eigenbasis can hide a transient growth of activity in the neurons
themselves.  Such growth can only be transient, for stability requirements
still demand that the activity decay asymptotically in time.

Purely nonnormal amplification that does not rely on dynamical slowing may be
ideally suited for sensory cortices that need to track inputs varying on fast
timescales (possibly of order $\tau$).  It has also been identified as a key
mechanism for short-term memory of past inputs, for in certain circumstances,
hidden feedforward dynamics enables the network to retain information about a
transient stimulus for a duration of order $N\tau$ \cite{Goldman09}. The
presence of noise as in Eq.\  \ref{ouprocess} could limit this memory duration
to $\sqrt{N} \tau$ \cite{Ganguli08,Ganguli09}, but this is still much longer
than the time $\tau$ in which individual neurons forget their inputs.

The above considerations apply to purposely structured networks
\cite{Ganguli08,Goldman09,Murphy09}.  It is not clear, however, how much of
this beneficial kind of amplification can be expected to arise in randomly
connected networks of excitatory and inhibitory neurons, a ubiquitous model of
cortical networks.  Murphy and Miller \cite{Murphy09} convincingly argued that
nonnormal amplification should generally be a key player in the dynamics of
balanced networks, i.e. when strong excitation interacts with equally strong
inhibition and when neurons can be only excitatory or inhibitory but not of a
mixed type.  When the connectivity is dense, or at least locally dense, weak
patterns of imbalance between excitation and inhibition can indeed be quickly
converted into patterns in which neurons of both types strongly deviate from
their mean firing rates.  Here, we revisit nonnormal amplification in the
context of \emph{random} balanced networks. We derive an analytical expression
for the purely nonnormal contribution to amplification in such networks. The
analysis reveals a strong tradeoff between amplification and dynamical slowing,
suggesting that the connectivity must be appropriately shaped for a network to
simultaneously exhibit fast dynamics and patterned amplification.

\section{Separating the effects of normal and nonnormal amplification}

\begin{figure}[tb!]
\centering
\includegraphics[scale=0.8]{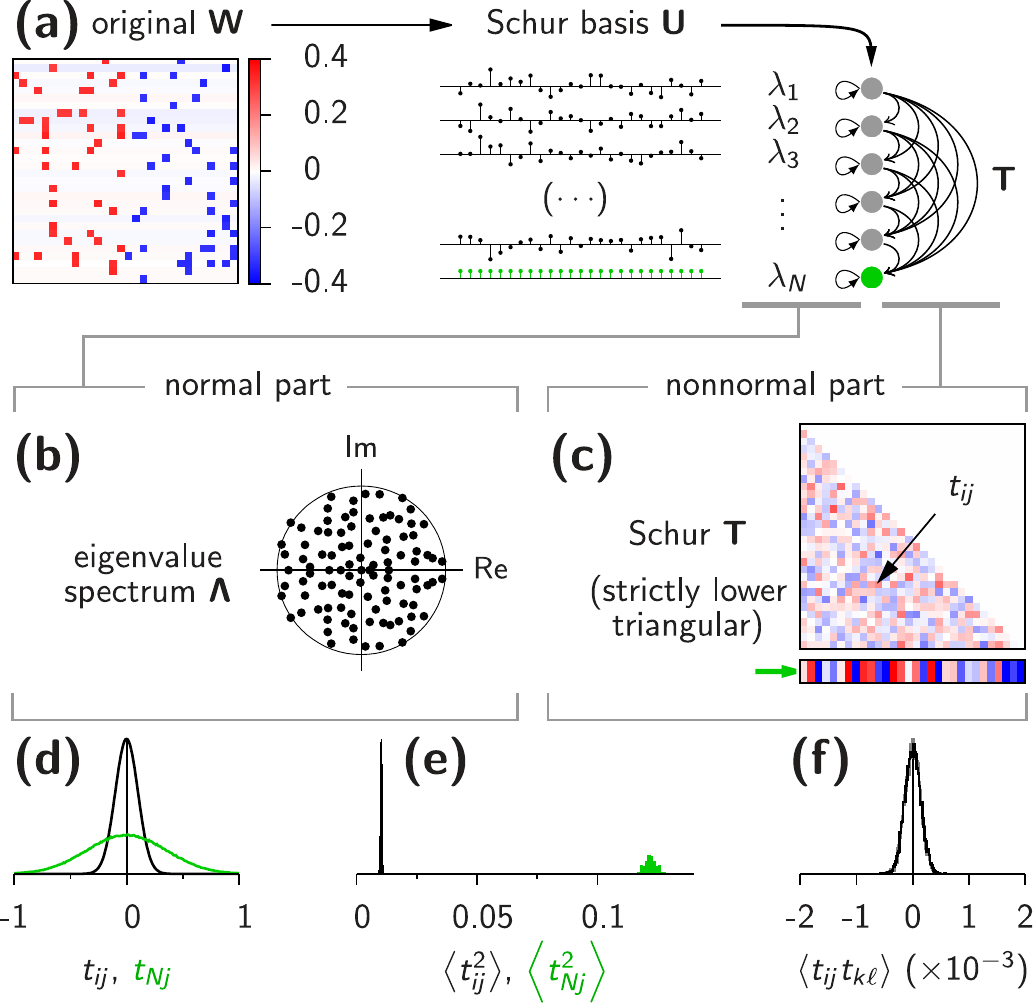}
\caption{\label{fig1}Teasing apart normal and nonnormal amplification in random
networks of excitatory and inhibitory neurons.  \textbf{(a)} Example sparse
neural connectivity matrix $\tmmathbf{W}$ (left, 50 exc. columns and 50 inh.
columns, thinned out to $30\times30$ for better visibility), a schematics of an
associated Schur basis $\tmmathbf{U}$ (center), and the corresponding abstract
network of Schur modes, in which the interactions are feedforward from top to
bottom (right).  The Schur vectors in $\tmmathbf{U}$ (center), orthogonal to
one another, represent patterns of neuronal activity in the original network.
The last Schur vector is explicitly chosen to be the uniform ``DC'' mode
$\tmmathbf{v}=(1,1,\cdots,1)/\sqrt{N}$ and is represented here in green.
\textbf{(b)} Amplification via dynamical slowing (``normal'' amplification) is
described by the set of eigenvalues
$\tmmathbf{\Lambda}=(\lambda_1,\ldots,\lambda_N)$ of $\tmmathbf{W}$, which for
a random network lies inside a disk centered around zero in the complex plane.
These eigenvalues determine the decay rates of the Schur patterns.
\textbf{(c)} Nonnormal amplification arises from the strictly lower-triangular
matrix $\tmmathbf{T}$ which describe the purely feedforward part of the
interactions between the Schur patterns.  The first non-zero entry in the upper
left corner of $\tmmathbf{T}$ is $t_{21}$ and represents the ``forward''
coupling from the first Schur mode onto the second.  The last row ($t_{N1},
t_{N2}, \ldots, t_{N(N-1)}$), zoomed-in (green arrow), is the coupling from the
first $N-1$ Schur modes onto the last (uniform) Schur mode $\tmmathbf{v}$.
\textbf{(d)} For a fixed large matrix \tmmathbf{W}, the non-zero entries
$t_{ij}$ in matrix $\mathbf{T}$ are approximately normally distributed with
zero mean and variance given by Eq.\ \ref{varrest} (black histogram, for
$j<i<N$). The entries in the last row have larger variance given by Eq.\
\ref{varfirstline} ($i=N$, green histogram).  \textbf{(e)} Moreover, the
variance $\left\langle t_{ij}^2 \right\rangle$ across many realisations of
$\tmmathbf{W}$ is the same for all $j<i<N$ (black histogram). Similarly,
$\left\langle t_{Nj}^2 \right\rangle$ is the same for all $j<N$ (green
histogram).  \textbf{(f)} The correlations $\left\langle t_{ij}
t_{k\ell}\right\rangle$ (for $i\neq k$ or $j\neq\ell$) are negligible, as seen
from a comparison of their empirical distribution (black) with surrogate data
from triangular matrices in which non-zero entries are all drawn i.i.d. (grey,
barely visible under the black curve).  The data for panels (d--f) was acquired
by Schur-transforming 5,000 random weight matrices of size $N=100$, drawn as
described in section \ref{link_schur_to_w} with connection density $p=0.1$ and
spectral radius $R=1$.  } \end{figure}

In the introduction, we have distinguished normal from nonnormal amplification.
The Schur decomposition (Fig.\ \ref{fig1}) -- a tool from linear algebra --
offers a direct way to assess the contributions of both mechanisms to the
overall amount of amplification $A(\tmmathbf{W})$. Any matrix $\tmmathbf{W}$
can be written as $\tmmathbf{U}^{\dag} \left(\tmmathbf{\Lambda} +
\tmmathbf{T}\right) \tmmathbf{U}$ where $\tmmathbf{U}=\{u_{ij}\}$ is unitary,
$\tmmathbf{\Lambda}$ is a diagonal matrix that contains the eigenvalues
$\lambda_k$ of $\tmmathbf{W}$, and $\tmmathbf{T}=\{t_{ij}\}$ is strictly
lower-triangular\footnote{Upper, not lower, -triangular $\tmmathbf{T}$ is more
common in the literature, but we prefer to keep the flow of information forward
(from the $1^\text{st}$ to the $N^\text{th}$ Schur mode) for notational
convenience in our calculations.} (Fig.\  \ref{fig1}\textbf{a--c}). The lines of
$\tmmathbf{U}$ are called the Schur vectors (or Schur modes) and are all
orthogonal to each other. If this decomposition is to avoid complex numbers,
$\tmmathbf{\Lambda}$ is only block-diagonal, with $2 \times 2$ blocks
containing the real and imaginary parts of complex conjugate pairs of
eigenvalues, and $1 \times 1$ blocks containing the real eigenvalues.
Importantly, because the Schur basis \tmtextbf{$\tmmathbf{U}$} is orthonormal,
the sum of variances in the basis of the Schur vectors is equal to the sum of
the single neuron activity variances.  Thus, in order to compute
$A(\tmmathbf{W})$, one can instead focus on the activity fluctuations in an
abstract network whose units correspond to spatial patterns of neuronal
activity (the Schur vectors) and interact with a connectivity matrix
$\tmmathbf{\Lambda}+\tmmathbf{T}$ (Fig.\
\ref{fig1}\textbf{a}, right). This matrix is lower-triangular, so the abstract
network is effectively feedforward. In the Schur network, unit $i$ receives its
input from all previous units $j<i$ according to the $i^\text{\tiny{th}}$ row
of $\tmmathbf{T}$. Without input, the activity of unit $i$ decays at a speed
governed by eigenvalue $\lambda_i$. 

A network with a \emph{normal} connectivity matrix would have only
self-feedbacks ($\tmmathbf{T}=0$), thus being equivalent to a set of
disconnected units with a variety of individual effective time constants,
reflecting dynamical slowing or acceleration. Amplification-by-slowing
therefore arises from $\tmmathbf{\Lambda}$ (Fig.\ \ref{fig1}\textbf{b}), which
summarises all the ``loopiness'' found in the original connectivity.
Conversely, when $\tmmathbf{\Lambda}=0$, all units share a common time constant
$\tau$ (which is also the time constant of the actual neurons) and interact in a
purely feedforward manner via matrix $\tmmathbf{T}$ (Fig.\
\ref{fig1}\textbf{c}). We refer to this case as ``purely nonnormal'', because
the network is then free of the unique dynamical consequence of normality,
namely a modification of the speed of the dynamics
\footnote{Note that quantifying nonnormality can be done in a
variety of ways, e.g. through several measures of ``departure from
normality'' \cite{Trefethen05}. Our concept of ``pure nonnormality'' is
therefore more specific to our particular purpose, in that it expresses the
absence of normal effects on the dynamics of the neurons.}.
``Purely nonnormal'' amplification therefore arises from matrix $\tmmathbf{T}$
that reveals the functional feedforward connectivity hidden in $\tmmathbf{W}$.

The latter situation ($\tmmathbf{\Lambda}=0$) is the focus of this paper. By
substituting $\tmmathbf{W}$ with $\tmmathbf{T}$ in Eq.\ \ref{ouprocess} and
subsequently calculating $A(\tmmathbf{T})$ as defined in Eq.\
\ref{amplificationdef}, we intend to reveal the fraction of the total
amplification $A(\tmmathbf{W})$ in the neuronal network that cannot be
attributed to dynamical slowing, but only to transient growth.  This
constitutes a functional measure of nonnormality.  We carry out this analysis
in a statistical sense, by calculating the expected amount of purely nonnormal
amplification $\left<A(\tmmathbf{T})\right>$ where the average
$\left<\cdot\right>$ is over the random matrix $\tmmathbf{W}$.  In section
\ref{link_schur_to_w}, the ensemble statistics of $\tmmathbf{W}$ are defined,
and related to the statistics of the non-zero entries of $\tmmathbf{T}$.  In
sections \ref{generalcase} and \ref{ampl_balanced_nets}, we perform the
calculation of $\left<A(\tmmathbf{T})\right>$.

\section{Schur representation of neural connectivity matrices}
\label{link_schur_to_w} Prior to calculating the nonnormal contribution to
amplification in realistic neural connectivity matrices, we first analyse the
statistical properties of the Schur triangle $\tmmathbf{T}$ derived from a
neuronal network where every
pair of neurons has a certain probability of being connected in either
direction.  Specifically, we consider networks of $N/2$ excitatory and $N/2$
inhibitory neurons, with connectivity matrices $\tmmathbf{W}$ drawn as follows
\footnote{It is straightforward to allow for any distribution
of non-zero weights; as it turns out, this Dirac delta distribution achieves
maximum nonnormal amplification.} (Fig.\ \ref{fig1}\textbf{a}):
\begin{equation}\label{wstats}
w_{ij} = \frac1{\sqrt{N}} \cdot \left\{
  \begin{array}{l}
  \left. \begin{array}{cc}
    +w_0 & \text{if } j \leq N/2 \\
    -w_0 & \text{if } j > N/2
  \end{array} \right] \text{ with proba. } p\\
0 \text{ with proba. } (1-p)
\end{array}
\right.
\end{equation}
Excitation and
inhibition are thus globally balanced. The $1/\sqrt{N}$ scaling ensures that in
the limit of large $N$, the eigenvalues $\left\{\lambda_k\right\}$ of
$\tmmathbf{W}$ become uniformly distributed inside the disk of radius
\begin{equation}\label{radius}
R \quad = \quad w_0 \sqrt{p(1-p)} 
\end{equation}
and centered around zero in the complex plane (Fig.\ \ref{fig1}\textbf{b}), with the
exception of a few outliers \cite{Rajan06}.  To push the outliers inside the
disk, we enforce that excitatory and inhibitory synapses cancel each other
precisely for each receiving neuron, i.e.  $\tmmathbf{Wv}=0$ with
$\tmmathbf{v}=(1,1,\cdots,1)/\sqrt{N}$ \cite{Rajan06,Tao11}. This constraint is
also essential to the identification of the ensemble statistics of
$\tmmathbf{T}$ as detailed below. Such a ``global balance'' can be achieved by
a Hebbian form of synaptic plasticity at inhibitory synapses in random spiking
networks \cite{Vogels11}. Here we enforce it by subtracting the row average (a
small number) from every row (which accounts for the barely visible horizontal
stripes in $\tmmathbf{W}$ of Fig.\ \ref{fig1}\textbf{a}).

The main point in relating the statistics of $\tmmathbf{T}$ to that of
$\tmmathbf{W}$ is to note that the Schur basis
is unitary, so that the sum of squares in $\tmmathbf{W}$ is also equal to the
sum of squares in $\tmmathbf{\Lambda}+\tmmathbf{T}$. Thus
\begin{equation}\label{sumsqrpreserved}
\sum_{1\leq i,j \leq N} w_{ij}^2 \quad=\quad \sum_{1\leq k\leq N} \left| \lambda_k \right|^2
+ \sum_{i>j} t_{ij}^2 
\end{equation}
From our choice of the weights $w_{ij}$ (Eq.\ \ref{wstats}) and assuming that $N$ is
large enough, we can derive $\sum w_{ij}^2 \simeq N p w_0^2$. Furthermore,
knowing that the eigenvalues lie uniformly inside the disk of radius $R$,
we can write $\sum \left| \lambda_k \right|^2 \simeq NR^2/2$
which is also valid for large $N$.
We replace these sums in Eq.\ \ref{sumsqrpreserved}, simplify the result using
Eq.\ \ref{radius}, and obtain the overall empirical variance of the non-zero
entries in $\tmmathbf{T}$, to leading order in $N$:
\begin{equation}\label{overalltvariance} 
\frac2{N(N-1)} \sum_{i>j} t_{ij}^2 \quad \simeq \quad \frac{R^2}N \cdot \frac{1+p}{1-p}
\end{equation}
Note that this empirical variance is not necessarily equal to the
\emph{ensemble} variance $\left< t_{ij}^2\right> - \left<t_{ij}\right>^2$ for
fixed $i$ and $j$. In fact, we have observed that if the non-unique Schur basis
is chosen arbitrarily, $\left< t_{ij}^2 \right>$ computed over many
realisations of $\tmmathbf{W}$ is not uniform across rows, but rather tends to
increase with row index $i$. This heterogeneity is difficult to characterise,
and undermines the calculation of amplification developed in the next section.
Fortunately, we can circumvent this problem by choosing the uniform eigenvector
$\tmmathbf{v}$ of $\tmmathbf{W}$ as the last Schur vector: $u_{Nk}=1/\sqrt{N}$
for all $k$%
\footnote{This is always possible, since a Schur basis can be constructed
through Gram-Schmidt orthonormalisation of the eigenbasis of $\tmmathbf{W}$, so
choosing $\tmmathbf{v}$ to enter the process first results in $\tmmathbf{v}$
being the last vector in a basis that makes $\tmmathbf{W}$
\emph{lower}-triangular}.
Coefficient $t_{ij}$ then becomes distributed with the same zero mean and
variance $\zeta^2$ for all $j<i<N$, with the exception of the $t_{Nj}$
coefficients which have higher variance $\zeta_0^2$ (black and green lines in
Fig.\  \ref{fig1}\textbf{d} and \ref{fig1}\textbf{e}, empirical observation).
Note also that the ensemble pairwise correlations between coupling strengths
$t_{ij}$ and $t_{k\ell}$ with $i\neq j$ or $j\neq \ell$ seem negligible (Fig.\
\ref{fig1}\textbf{f}).

\begin{figure}[tb!]
\centering
\includegraphics[scale=0.8]{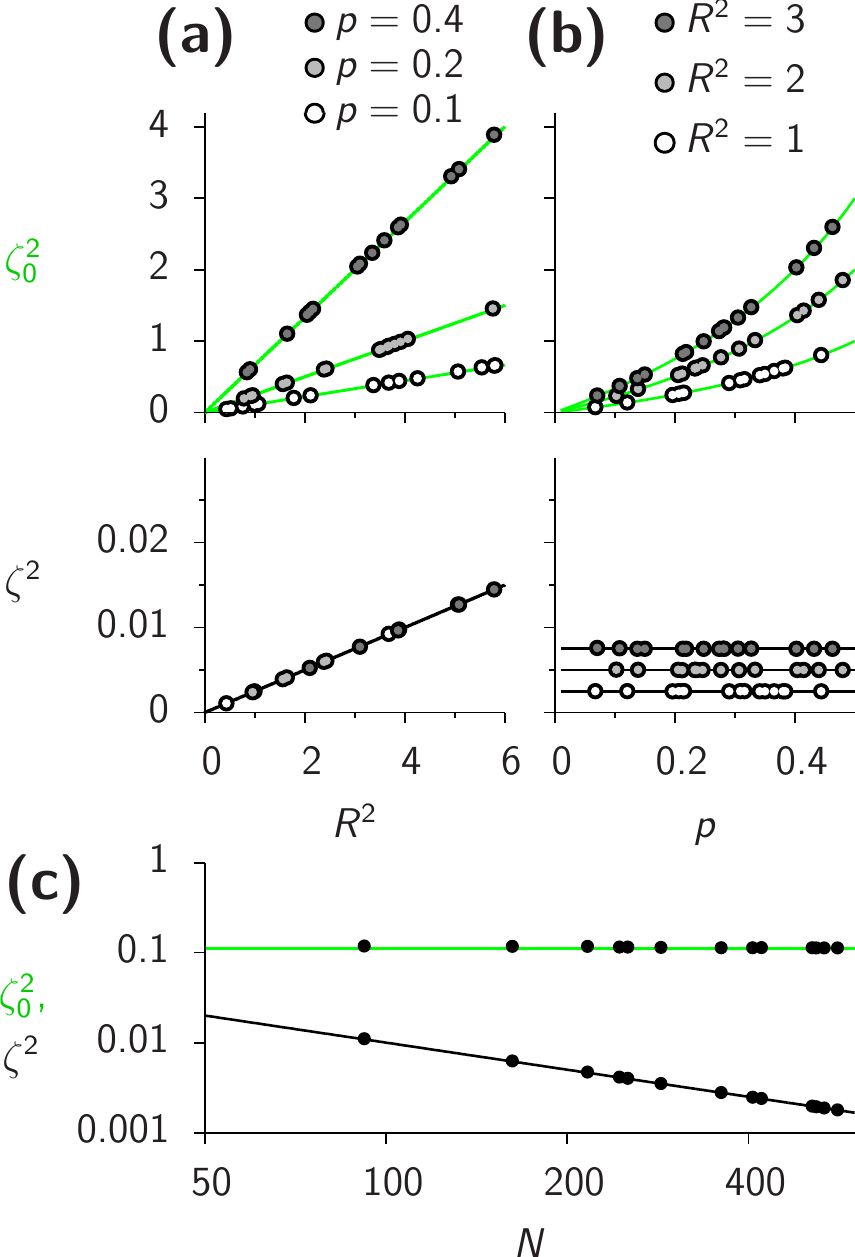}
\caption{\label{fig2}Linking the Schur triangle to the parameters of the neural
connectivity matrix.
\textbf{(a)} The variance of the entries in the strict lower triangle
$\mathbf{T}$ scales linearly with the square of the spectral radius $R^2$ of
the original weight matrix $\mathbf{W}$.  For the last row of $\mathbf{T}$, the
slope of $\zeta_0^2$ depends on the connection probability $p$ (top plot). For the
rest of $\mathbf{T}$, $\zeta^2$ depends only on $R^2$ (bottom plot).  Each
point was obtained by empirically estimating $\zeta^2$ and $\zeta_0^2$ from 10
different Schur-transformed random neural weight matrices of size $N=400$.
Lines denote the analytical expressions in Eqs.\  \ref{varfirstline} and
\ref{varrest}.  
\textbf{(b)} $\zeta_0^2$ in the last row of $\mathbf{T}$ scales super-linearly
with the connection density $p$ (top plot). In contrast, $\zeta^2$ does not
depend on $p$ (bottom plot).  
\textbf{(c)} In the last row of $\mathbf{T}$, the variance is network
size-independent (green line). In the rest of $\mathbf{T}$, the variance is
inversely proportional to $N$ (black line, note the log-log scale).  }
\end{figure}

We now proceed in two steps. First, we focus on the variance of the elements in
the \emph{last} row of the Schur matrix $\tmmathbf{T}$,
and then we turn to all the other non-zero components.
To calculate variance $\zeta_0^2 = \left\langle t_{Nj}^2 \right\rangle$ we use
the definition of $\tmmathbf{T}$ and write for $j<N$ 
\begin{equation}\label{abnormal_calc} \begin{split}
t_{Nj} &\quad=\quad \sum_{\ell = 1}^N \sum_{k = 1}^N u_{N k} w_{k \ell} u_{j \ell} \\
 &\quad=\quad \frac{1}{\sqrt{N}} \sum_{\ell = 1}^N \left( \sum_{k = 1}^N w_{k \ell}
   \right) u_{j\ell} 
\end{split}
\end{equation}
To leading order in $N$ we can write $\sum_k w_{k\ell} = \pm p w_0 \sqrt{N}$
where the $\pm$ sign depends on $\ell$ being smaller than $N/2$ ($+$,
excitatory) or greater ($-$, inhibitory) -- see Fig.\ \ref{fig1}\textbf{a}.  
For $j<N$, the $j^\text{\tiny{th}}$ Schur vector $\tmmathbf{U}_j$ is orthogonal
to the last Schur vector $\tmmathbf{v}\propto(1,1,\ldots,1)$, so its components
strictly sum to zero: $\sum_\ell u_{j\ell} = 0$.
Moreover, because of the normalization, 
$\sum_\ell u_{j\ell}^2 = 1$.
We can therefore approximate $u_{j\ell}$ by a stochastic process with
zero mean and variance $1/N$.
Assuming the $u_{j\ell}$ are uncorrelated, the variance of $t_{Nj}$
is thus simply $ w_0^2 p^2 $ to leading order, which according to Eq.\ \ref{radius}
is also
\begin{equation}\label{varfirstline}
\left<t_{Nj}^2\right> \quad\equiv\quad \zeta_0^2 \quad=\quad \frac{R^2 p}{1-p} 
\end{equation}
Notably, the variance $\zeta_0^2$ in the last row of coupling matrix
$\tmmathbf{T}$ is of order 1, and depends super-linearly on the connectivity
density $p$ (Fig.\ \ref{fig2}, green lines).

We now turn to the other rows $i<N$ of the Schur matrix $\tmmathbf{T}$.
Because all components $t_{ij}$ for $j<i<N$ seem to come from the same
distribution and look uncorrelated (Fig.\ \ref{fig1}\textbf{d--f}),
the empirical estimate of their variance $2\sum_{j<i<N} t_{ij}^2/(N-1)(N-2)$
coincides with the ensemble variance $\zeta^2 \equiv \left<t_{Nj}^2\right>$ so
long as $N$ is large enough.  Similarly, we can write $\sum_j t_{Nj}^2 / (N-1)
= \zeta_0^2$.
Thus, the l.h.s. of Eq.\ \ref{overalltvariance} becomes
$\zeta^2 + 2\zeta_0^2/N$ to leading order in $N$.
Using Eqs.\ \ref{overalltvariance} and \ref{varfirstline} we conclude 
\begin{equation}\label{varrest}
\left<t_{ij}^2\right> \quad\equiv\quad \zeta^2 \quad=\quad\frac{R^2}{N}
\end{equation}
Figure \ref{fig2} shows that Eqs.\ \ref{varfirstline} and \ref{varrest}
provide a good match to numerical results.

At this point we can already draw a few conclusions.
Suppose each unit in our Schur network receives external input of variance $1$.
First, since the uniform mode
$\tmmathbf{v}$ receives network input from the remaining $N-1$ Schur patterns
with coupling coefficients of order 1 (Eq.\ \ref{varfirstline}), 
we expect the global (``DC'') population
activity $\tmmathbf{x}\cdot \tmmathbf{v}$ to fluctuate macroscopically, i.e.
with a variance of order $N$. In contrast, the rest of the Schur modes should
display fluctuations of order 1. Second, we directly see that making the
network denser (i.e. increasing $p$) can only result in larger DC fluctuations,
but no further amplification of the other (zero-mean) Schur patterns. This is
because $\zeta_0^2$, but not $\zeta^2$, depends on $p$.  Third, it is easy to
see where these large DC fluctuations would originate from.
Imagine breaking the overall exc.-inh. balance in the network activity by a
small amount, e.g. by initialising the network state $\tmmathbf{x}$ to
$\tmmathbf{d}=(1,\ldots,1,-1,\ldots,-1)/\sqrt{N}$, where we emphasize the
scaling in $1/\sqrt{N}$.  According to Eq.\ \ref{ouprocess}, the transient
response to this perturbation is roughly $\tmmathbf{W}\tmmathbf{d}$, which to
leading order in $N$ equals 
\begin{equation}\label{dcorigin}
\tmmathbf{W}\tmmathbf{d} \quad\simeq\quad pw_0 (1,1,\cdots,1)
\end{equation}
We note that the $1/\sqrt{N}$ scaling is gone. Thus, the network responds to a
microscopic global balance disruption -- a state in which the deviation between
the excitatory and inhibitory population firing rates is of order $1/\sqrt{N}$
-- by an excursion of order 1 in the combined firing rate of both populations
(see \cite{Murphy09} for a more in-depth discussion of this effect).  Finally,
it is instructive to see what happens when the functional feedforward link
from $\tmmathbf{d}$ to $\sqrt{N}\cdot \tmmathbf{v}$ -- expressed in Eq.\
\ref{dcorigin} -- is removed from $\tmmathbf{W}$. This can be achieved by
transforming $\tmmathbf{W}$ into $\tmmathbf{W}'$ given by 
\begin{equation}
\label{demean} \tmmathbf{W}'=\tmmathbf{W}-\frac{p w_0}{\sqrt{N}}
(1,\cdots,1)^\dag
(1,\cdots,1,-1,\cdots,-1) \end{equation} It is easy to see that $\tmmathbf{W}'
\tmmathbf{d} = 0$.  In this case, calculations similar to Eqs.\
\ref{sumsqrpreserved}--\ref{varfirstline} yield $\zeta_0^2 = \zeta^2 =
R^2/N$ so that the DC fluctuations are back to order $1$: the amplification
along the DC mode becomes comparable in magnitude to the amplification that
occurs along any other Schur directions.  Note that the operation in Eq.\
\ref{demean} effectively shifts the mean excitatory (resp.  inhibitory) weight
from $pw_0/\sqrt{N}$ (resp.  $-pw_0/\sqrt{N}$) to zero.  We now substantiate
these preliminary conclusions through a direct calculation of nonnormal
amplification.

\section{Amplification in random strictly triangular networks}
\label{generalcase}

We have seen in the preceding two sections that a randomly coupled network of
excitatory and inhibitory neurons can be transformed via a unitary Schur basis
into a different network where the couplings between units are given by a lower
triangular matrix (Fig.\ \ref{fig1}\textbf{a}). Furthermore, the ``purely nonnormal'' part of the
amplification of the external noisy input in the original network of neurons
corresponds to the activity fluctuations in the new feedforward network where
all self-couplings are neglected (Fig.\ \ref{fig1}\textbf{c}). Finally, we have
also seen that it is possible to constrain the Schur basis such that the
couplings between the first $N-1$ units in the feedforward network are
independently distributed with the same zero-mean and a variance given by the
parameters of the original synaptic weights (Eq.\ \ref{varrest}). In this
section, we therefore study this ``canonical'' case, starting directly from a
strictly lower-triangular matrix $\tmmathbf{T}$ and ignoring -- for the moment
-- the transformation that gave rise to $\tmmathbf{T}$.  

\begin{figure}[tb!]
\centering
\includegraphics[scale=0.8]{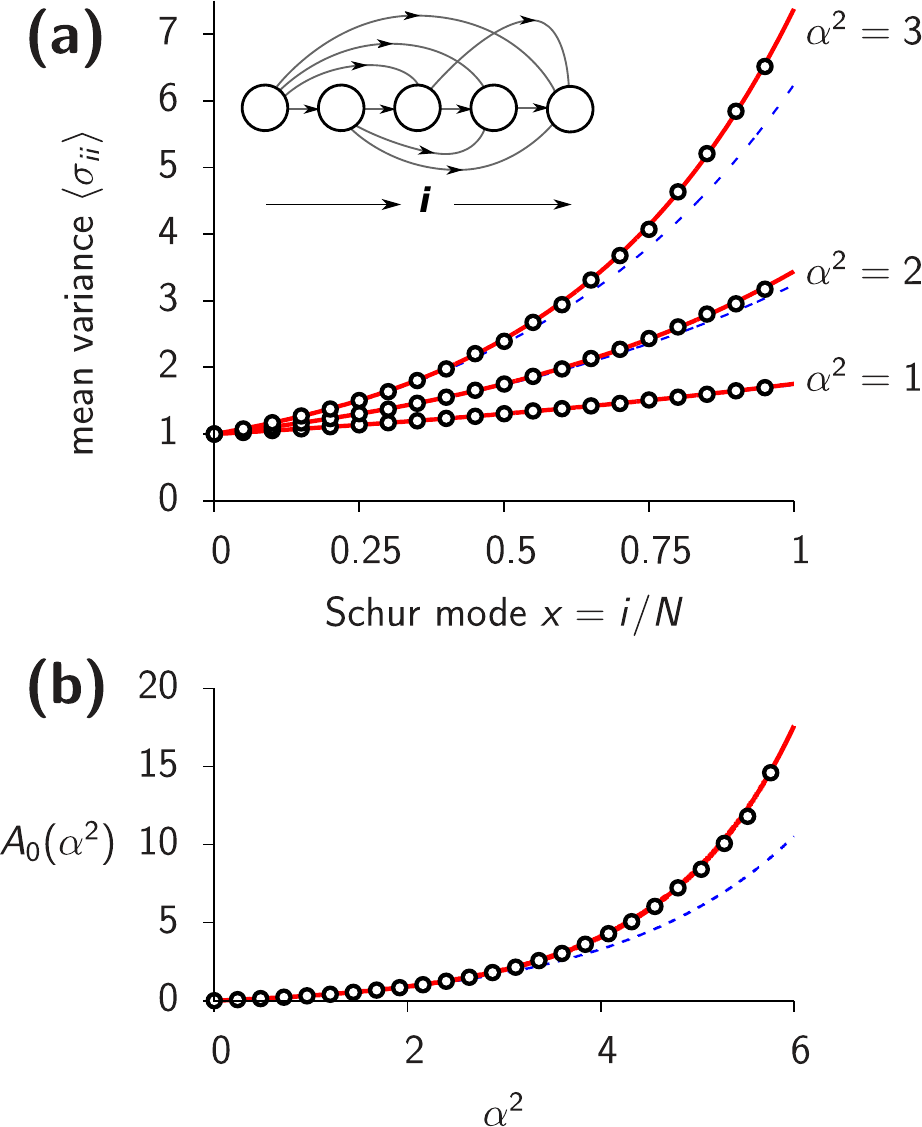}
\caption{\label{fig3} Analytical result for a feedforward network of $N$
Ornstein-Uhlenbeck processes coupled via a random strictly lower-triangular
matrix (inset).
\textbf{(a)} The expected activity variance $\left<\sigma_{ii}\right>$
accumulates super-linearly from the first unit to the last down the feedforward
chain. Dashed blue lines depict the closed-form lower-bound of Eq.\
\ref{secondorderprofile}. Red lines denote the exact solution given in Eq.\
\ref{powerseries}, truncated to $K=10$. Open circles represent the numerical
solution of Eq.\ \ref{ouprocess} -- or more exactly, the numerical solution of
equation $\ref{sigma}$ given in the appendix -- averaged over 20 randomly
generated matrices of size $N=500$. Each matrix $\tmmathbf{T}$ is characterised
by the variance $\alpha^2/N$ of the coupling coefficients $t_{ij}$ with $j<i$.
The strength of the external noise driving each unit independently
is set to $\sigma_\xi^2 = 2/\tau$ so that all activity variances in the network
would be 1 should the couplings $t_{ij}$ be set to 0.
\textbf{(b)} The total amplification (the area under the curves in (a), minus 1)
explodes with increasing variance $\alpha^2/N$ in the triangular connectivity matrix.
Points and lines have the same meaning as in (a).}
\end{figure}

We want to solve for the expected variances of $N~\gg~1$ Ornstein-Uhlenbeck
processes (as in Eq.\  \ref{ouprocess}) coupled by a strictly lower-triangular
weight matrix $\tmmathbf{T}$ (therefore describing a purely feedforward
network, see inset in Fig.\  \ref{fig3}\textbf{a}).  We assume all non-zero coupling
strengths to be sampled i.i.d. from some common distribution with zero mean and
variance $\alpha^2/N$. Due to the coupling matrix, the fluctuations that the
external input causes in the first unit feed and augment those it causes in
unit 2. The third unit in turn fluctuates due to the external input and the
activities of units 1 and 2, and so on. We therefore expect the activity
variance $\sigma_{ii}$ in unit $i$ to increase with index $i$.  In appendix
\ref{appendix}, we show that in the limit of large $N$ and for some fixed
$0\leq x \leq 1$, the relative expected variance of the activity in unit $i=xN$
is $g(i/N) \equiv 2 \left<\sigma_{ii}\right> / \tau \sigma_\xi^2$ where the
function $g(x)$ is lower-bounded in closed form by
\begin{equation}\label{secondorderprofile}
  \begin{split}
  g^{\text{\tiny{LB}}} \left( x \right) &\quad=\quad \frac{1}{3 + \sqrt{3}} \exp \left( \frac{1 -
  \sqrt{3}}{4} \alpha^2 x \right) \\
&\quad + \quad \frac{2 + \sqrt{3}}{3 + \sqrt{3}}
  \exp \left( \frac{1 + \sqrt{3}}{4} \alpha^2 x \right)
\end{split}
\end{equation}
(Fig.\ \ref{fig3}, dashed blue curves). We also derive the exact solution as a
power series
\begin{equation}
\label{powerseries}
g(x)\quad=\quad\lim_{K\to\infty}\sum_{k=0}^K \beta_k x^k
\end{equation}
with the $\beta_k$ coefficients defined recursively as
\begin{equation}\label{betacoeffs}
\begin{split}
\beta_0 &= 1 \\
\beta_k &= \frac{\alpha^2}{2k!} \sum_{\ell=0}^{k-1} \frac{(2\ell)!\, (k-\ell-1)!}{\ell!\, (\ell+1)!}
 \left(\frac{\alpha^2}4\right)^\ell \, \beta_{k-\ell-1}
\end{split}
\end{equation}
The overall amplification $A_0(\alpha^2)$ in the network is subsequently
obtained by integrating this variance profile $g(x)$ from $0$ to $1$, which
corresponds to taking Eq.\ \ref{amplificationdef} to its $N\to\infty$ limit:
\begin{equation}
\label{finalamp}
A_0(\alpha^2) \quad=\quad \left( \lim_{K\to\infty}\sum_{k=0}^{K} \frac{\beta_k}{k+1} \right) - 1
\end{equation}
Figure \ref{fig3} shows that Eqs.\ \ref{powerseries} and \ref{finalamp}
indeed converge to the empirical mean variance profile and mean amplification
as the cut-off parameter $K$ of the power series becomes large  (red lines,
$K=10$). Figure \ref{fig3}\textbf{b} furthermore shows how amplification explodes
with the variance $\alpha^2/N$ of the feedforward couplings in the network.

\section{Amplification in random balanced networks} 
\label{ampl_balanced_nets}

Using the canonical result of the previous section that is restricted to
homogeneous random lower-triangular matrices, we now calculate $A(R,p) \equiv
\left\langle A(\tmmathbf{T}) \right\rangle$ with $\tmmathbf{T}$ originating
from the Schur decomposition of a neuronal connectivity matrix as in section
\ref{link_schur_to_w}, with connection density $p$ and spectral radius $R$.
Equation \ref{powerseries} can directly be applied with $\alpha^2/N = \zeta^2 =
R^2/N$ (see Eq.\ \ref{varrest}) to describe the activity fluctuations of the
first $N-1$ Schur modes. The last Schur unit, however, receives feedforward
input with couplings of variance $\zeta_0^2 \neq \zeta^2$ (Eq.\
\ref{varfirstline}). Consequently, the expected variance
$\left<\sigma_{NN}\right>$ of its temporal fluctuations has to be treated
separately. In appendix \ref{appendixuniformmode}, we show that
\begin{equation}
\label{dc}
\lim_{N \rightarrow \infty} \frac{\left<\sigma_{NN}\right>}{N} \quad=\quad
 \frac{\sigma_\xi^2 \tau}2 \cdot \frac{p}{1-p}
   \left[ g \left( 1 \right) - 1 \right] 
\end{equation}
where $g$ is given by Eqs.\ \ref{powerseries} and \ref{betacoeffs}, here with
$\alpha=R$. Gathering the contributions of all Schur modes, we obtain the
expected overall amount of purely nonnormal amplification in $\tmmathbf{W}$: 
\begin{equation}\label{finalsecondorder}
  A(R,p) \quad =\quad A_0(R^2) + 
\frac{p}{1 - p} \left[ g \left( 1 \right) - 1 \right]
\end{equation}
with $A_0(R^2)$ given by Eq.\ \ref{finalamp}.

\begin{figure}[tb!]
\centering
\includegraphics[scale=0.8]{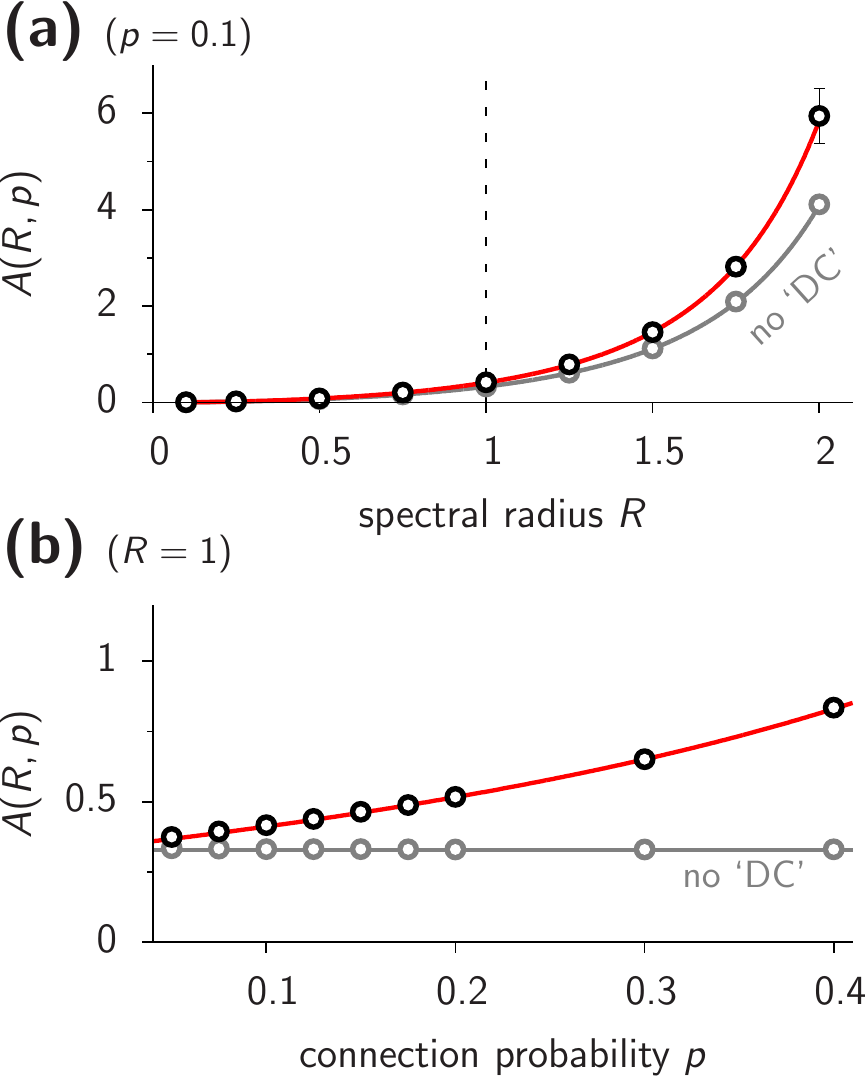}
\caption{\label{fig4}Nonnormal amplification in random neuronal networks.
\textbf{(a)} The mean amount of purely nonnormal amplification $\left\langle
A(\mathbf{T}) \right\rangle \equiv A(R,p)$ is reported as a function of the
spectral radius $R$ of $\mathbf{W}$.  Open circles denote the numerical
solution of Eq.\ \ref{sigma} averaged over 20 randomly drawn connectivity
matrices with connection density $p=0.1$ and size $N=500$.  Errorbars denote
the standard deviation over all trials.  Red lines depict the exact solution in
Eq.\ \ref{finalsecondorder}. The grey line and grey circles indicate the mean
removal of Eq.\ \ref{demean} applied to $\tmmathbf{W}$, which effectively
removes the global macroscopic fluctuations of the entire population (labelled
``no DC''). The dashed vertical line represents the limit of linear stability,
beyond which the nonnormal part of amplification is still well-defined.
\textbf{(b)} Same as in (a), now as a function of the connection density $p$
for a fixed $R=1$. In both (a) and (b), parameters $p$ and $R$ fully determined
the value $\pm w_0/\sqrt{N}$ of the nonzero synaptic weights as
$w_0 = R/\sqrt{p(1-p)}$ (cf. Eq.\ \ref{radius}).
}
\end{figure}

Figure \ref{fig4}\textbf{a} shows that the nonnormal contribution to
amplification in the neuronal network explodes with the spectral radius $R$ of the
connectivity matrix $\tmmathbf{W}$. This is because the amplification of the 
first $N-1$ Schur units explodes with the variance $\zeta^2$ of their feedforward interactions
(Fig.\ \ref{fig3}\textbf{b}) and that $\zeta^2$ is directly related to $R$ (Eq.\ \ref{varrest}).
Note that for $R>1$ (to the right of the dashed vertical line in Fig.\
\ref{fig4}\textbf{a}), the network of neurons is unstable.  Although the
concept of amplification in an unstable network is ill-defined, the ``purely
nonnormal'' part of the total (infinite) amplification remains bounded.
Indeed, the purely feedforward network $\tmmathbf{T}$ derived from the Schur
decomposition of $\tmmathbf{W}$ is itself always stable, since zero is the only
eigenvalue of $\tmmathbf{T}$.  The instability in $\tmmathbf{W}$ arises from
purely normal effects, when the real part of one eigenvalue of $\tmmathbf{W}$
exceeds unity so that dynamical slowing becomes infinite.

Equation \ref{dc} confirms what we had previously discussed at the end of
section \ref{link_schur_to_w}: the last Schur unit has temporal fluctuations
$\tmmathbf{v}\cdot\tmmathbf{x}(t)$ of variance $\mathcal{O}(N)$. Those fluctuations
thus make up for a finite fraction of the total nonnormal amplification (the last
term in Eq.\  \ref{finalsecondorder}) as $N\to\infty$. 
Because the last Schur vector is the normalised uniform spatial pattern
$(1,\ldots,1)/\sqrt{N}$, the variance of the overall population activity $\mu(t) \equiv
\sum x_i(t) /N = \sqrt{N} (\tmmathbf{x}\cdot\tmmathbf{v}(t))$ is of order $1$.
As we had foreseen in section \ref{link_schur_to_w}, one can restore the $1/N$
scaling of the these ``DC'' fluctuations $\left<\mu^2(t)\right>$ by performing
the operation of Eq.\ \ref{demean} on the connectivity matrix $\tmmathbf{W}$,
i.e. subtracting a common constant from all excitatory weights (including zero
weights) to make sure that they average to zero, and adding the same constant
to all inhibitory weights with the same purpose. This situation is depicted by
the grey curves in Fig.\ \ref{fig4}. Figure \ref{fig4}\textbf{b} shows that only
these DC fluctuations depend on the connectivity density $p$.

Overall, Fig.\ \ref{fig4}\textbf{a} allows us to draw two important
conclusions. On the one hand, if the level of dynamical slowing is to be kept
low ($R \ll 1$), only modest levels of amplification can be achieved (see the
small amount of nonnormal amplification on the l.h.s. of the dashed vertical
line). For example, if no mode is to decay with more than twice the single
neuron time constant ($\text{Re}(\lambda)<1/2$), the average variance cannot
exceed that of a disconnected network by more than 10\%.  On the other hand,
the nonnormal contribution to amplification explodes with increasing $R$, i.e.
with increasing synaptic strengths if the connection density is taken fixed.
This suggests that strong transient amplification without dynamical slowing can
only be achieved in structured, ``less random'' networks. The structure must
allow the synaptic couplings to assume larger values without causing the
eigenvalue spectrum of $\tmmathbf{W}$ to reach instability.

\section{Different numbers of excitatory and inhibitory neurons}

\begin{figure}[tb!]
\centering
\includegraphics[scale=0.93]{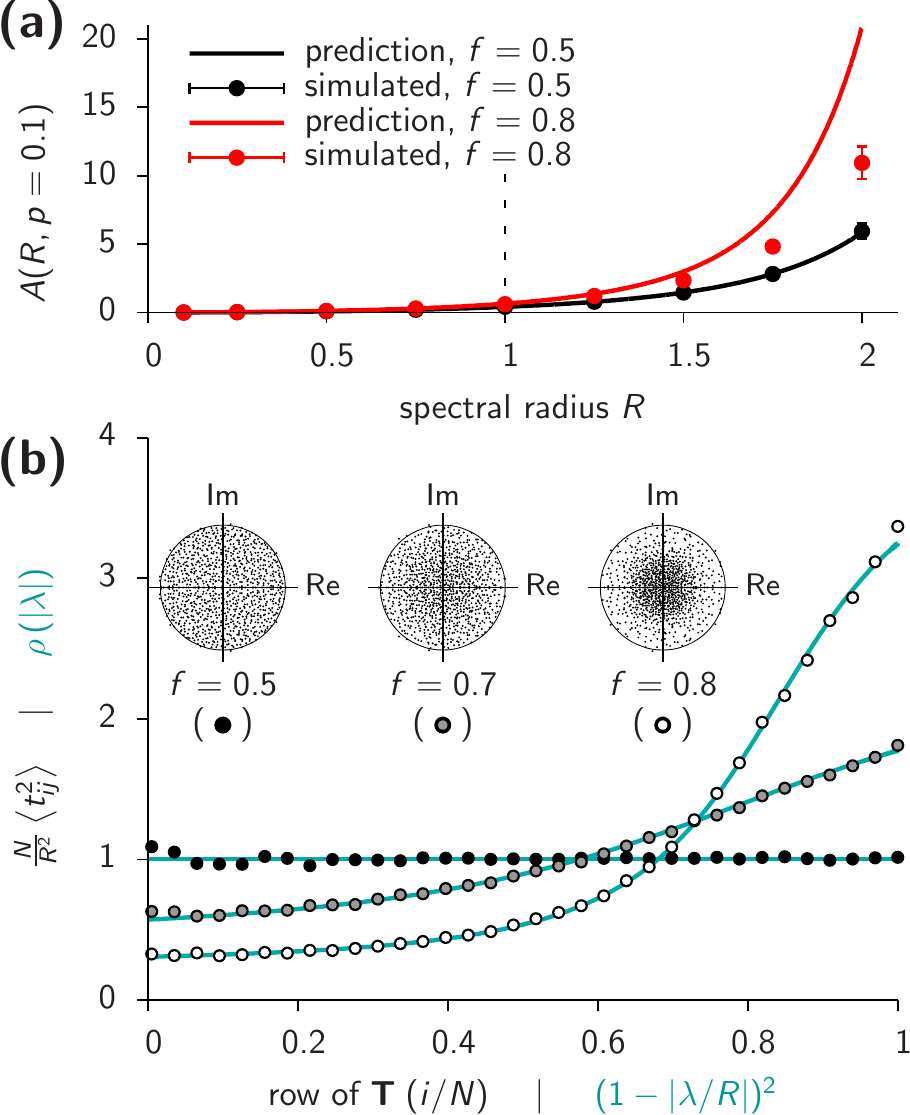}
\caption{\label{fig5}Networks with different numbers of excitatory and
inhibitory neurons.  
\textbf{(a)} Nonnormal amplification as a function of the spectral radius
$R$ of $\tmmathbf{W}$, in sparse random balanced networks with $fN$ excitatory
and $(1-f)N$ inhibitory neurons, for $f=0.5$ (black) and $f=0.8$ (red). The
connection density $p$ was set to $0.1$. The dashed vertical line
represents the limit of linear stability, beyond which the nonnormal part of
amplification is still well-defined. Solid circles were obtained by averaging the
numerical solution of Eq.\ \ref{sigma} for 20 random matrices of size $N=500$.  Errorbars
denote standard deviation over all trials.
\textbf{(b)} Filled circles show the scaled variance $N\left< t_{ij}^2
\right>/R^2$ of the non-zero Schur couplings in row $i$ as a function of $i/N$ and
for three different values of $f$. These variances were computed by
Schur-transforming 100 matrices of size $N=200$, with $R=1$ and $p=0.1$.  Cyan
lines denote the density $\rho$ of eigenvalues $\lambda$ inside the unit disk
\cite{Rajan06}, as a function of $(1-\left|\lambda/R\right|)^2$.  Insets show
the eigenvalue spectra of three example matrices of size $N=1000$.  }
\end{figure}

We now consider the biologically more plausible case of different numbers of
excitatory and inhibitory neurons. Typical models of cortex assume $fN$
excitatory neurons and $(1-f)N$ inhibitory neurons with $f=0.8$ or similar.  In
this case, the eigenvalues $\lambda$ are no longer uniformly scattered inside
the disk of radius $R$ in the complex plane\footnote{Rajan and Abbott showed
that this happens when the \emph{variances} of the excitatory and inhibitory
weights differ (the variances comprise both the zero and non-zero synapses).
Decreasing the number of inhibitory neurons in a balanced network requires the
strength of inhibition to be increased. In sparse networks like ours, this
automatically makes the overall variance of the inhibitory synapses larger than
that of excitatory synapses, hence the observed effect on the eigenspectrum.},
but become more concentrated in the middle following a radially symmetric density
$\rho\left(\left|\lambda\right|\right)$ known analytically from \cite{Rajan06}
(Fig.\ \ref{fig5}\textbf{b}, insets).
As before, we consider the case where excitatory (resp.  inhibitory) synaptic
couplings are $0$ with probability $(1-p)$, and $+w_E/\sqrt{N}$ (resp.
$-w_I/\sqrt{N}$) otherwise.  The global balance condition reads $f w_E = (1-f)
w_I$.
To impose a given spectral radius $R$, we set $w_E^2 = w_0^2 (1-f)/f$ and
$w_I^2 = w_0^2 f/(1-f)$ with $w_0^2 = R^2 / p(1-p)$. 

The results of section \ref{link_schur_to_w} regarding the variances in the
Schur triangle have to be adjusted to accommodate these modifications.  The
derivation of $\zeta_0^2$ is left unchanged, so that the couplings $t_{Nj}$
onto the uniform mode $\tmmathbf{v}$ still have the variance given by Eq.\
\ref{varfirstline}, which notably does not depend on $f$. Using Eq.\
\ref{sumsqrpreserved}, we can then write down the empirical variance in the
first $N-1$ rows of $\tmmathbf{T}$ as  
\begin{equation}\label{overalltvariance_f}
\frac2{N(N-1)} \sum_{j<i<N} t_{ij}^2 = \frac2N \left(R^2 - \int_0^{R} r \rho(r) dr \right)
\end{equation}
Unfortunately, the \emph{ensemble} variance $\left<t_{ij}^2\right>$ for fixed
$i$ and $j$ is in general different from the average across matrix elements
given by Eq.\ (\ref{overalltvariance_f}). Indeed,
contrary to the case $f=0.5$ considered in 
section \ref{link_schur_to_w}, the non-zero elements of $\tmmathbf{T}$ \emph{no
longer} have the same ensemble variance. Instead, $\left\langle t_{ij}^2
\right\rangle$ grows with row index $i<N$, and this profile interestingly
matches the density of eigenvalues $\rho$ \footnote{This happens provided the
eigenvectors are sorted in decreasing order of their corresponding eigenvalue
moduli, prior to going through the Gram-Schmidt orthonormalisation process.
This results in a unique Schur basis.}, according to
\begin{equation}\label{surprising_identity}
\frac{N}{R^2}\left\langle t_{ij}^2 \right\rangle \quad =\quad \rho\left[R\left(1 - \sqrt{\frac{i}N}\right) \right]
\quad \text{for} \quad j<i<N 
\end{equation}
This is depicted in Fig.\ \ref{fig5}\textbf{b}.  

In a feedforward network like that of Schur units considered here, a good
strategy to generate greater amplification would be to give comparatively more
power to the couplings onto earlier nodes. This is because amplification builds
up superlinearly along the feedforward chain (Fig.\ \ref{fig3}), so that
boosting early nodes exacerbates the avalanche effect (see also
\cite{Ganguli08}).  Setting $f$ to more than $0.5$ does precisely the contrary:
couplings onto early nodes become comparatively smaller in magnitude, as shown by
the filled circles in Fig.\  \ref{fig5}\textbf{b}. Therefore, simply replacing
$\alpha^2/N$ in Eq.\  \ref{finalamp} by the empirical variance of Eq.\
\ref{overalltvariance_f} yields an over-estimation of the true amplification in
the first $N-1$ Schur units (compare the red line with the red circles in Fig.\
\ref{fig5}\textbf{a}).  We found it difficult to incorporate this variance
profile $\left\langle t_{ij}^2\right\rangle$ into the derivation of appendix
\ref{appendix}, so we can only consider as accurate the results of numerical
simulations.

The conclusions reached at the end of section \ref{ampl_balanced_nets} do not
change significantly under the more realistic assumption of $f=0.8$.  Although
amplification almost doubles relative to $f=0.5$, it remains very weak in the
stable regime (to the left of the dashed vertical line in Fig.\
\ref{fig5}\textbf{a}), confirming that amplification can only come with
substantial dynamical slowing when connections are drawn at random.

\section{Example of network structure for nonnormal amplification}

\begin{figure}[tb!]
\centering
\includegraphics[scale=0.95]{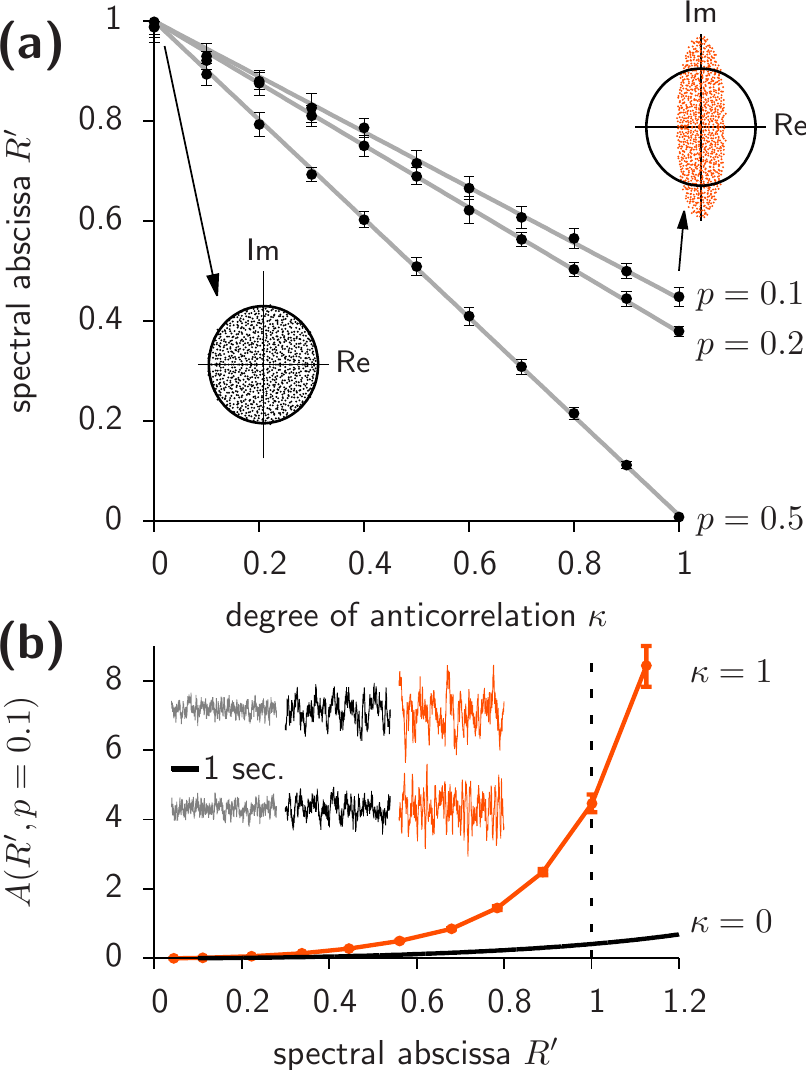}
\caption{\label{fig6}Example of a network structure that favors nonnormal
amplification: unidirectional vs. bidirectional synaptic connections.
\textbf{(a)} We varied the degree of anticorrelation between reciprocal weights
in the connectivity matrix, as the fraction $\kappa$ of the maximum value it
can assume, which is dictated by the connection sparsity (see
text). This caused the eigenspectrum to stretch more and more along the
vertical axis (compare the two insets), effectively decreasing the spectral
abscissa $R^\prime$ (black filled circles).  Empirical data was obtained from
numerically computing the eigenvalues of 20 different matrices of size $N=500$.
Errorbars denote standard deviations over all trials. Gray lines are linear
fits. \textbf{(b)} Nonnormal amplification as a function of the spectral abscissa
$R'$. When all connections between an excitatory (E) and an inhibitory (I) cell
are made reciprocal, while all E~$\to$~E and I~$\to$~I connections are kept
unidirectional (orange curve, corresponding to $\kappa=1$ in (a)), stronger
amplification is obtained in the fast dynamical regime ($R'\ll 1$). The
black curve is here reproduced from Fig.\ \ref{fig4} (purely random case,
$\kappa=0$) for comparison.  The inset displays examples of 4-second snapshots
of activity in a disconnected network (grey), a random network (black,
$\kappa=0$), and a maximally (though not fully) antisymmetric network (orange,
$\kappa=1$).  The spectral abscissa was set to $R'=0.9$. Traces were obtained
from a direct simulation of Eq.\  \ref{ouprocess}, and are shown here only for
two randomly chosen neurons.} \end{figure}

Here we show that random networks can be minimally structured in such a way
that strong nonnormal amplification occurs already in the fast dynamical
regime. We exploit the fact that correlations in the connectivity matrix can
modify the shape of the eigenvalue spectrum.  Symmetrising (or
anti-symmetrising) $\tmmathbf{W}$ has been shown to generate elliptical (as
opposed to circular) eigenspectra, in the case of ``centered'' matrices where
the distinction between excitatory and inhibitory neurons is not made
\cite{Sommers88}. Here we consider a modification of the sparse neural matrices
studied in section \ref{link_schur_to_w} that achieves this slimming effect in
the case of balanced networks (see the insets in Fig.\  \ref{fig6}\textbf{a}).
All non-zero entries assume a value $\pm w_0/\sqrt{N}$, the sign depending on
the excitatory versus inhibitory nature of the presynaptic neuron. Whether a
connection exists (non-zero entry) is decided as follows. Connection $w_{ij}$
with $i\geq j$ exists with probability $p$. If $i \neq j$, the reciprocal
connection $w_{ji}$ then exists with probability $p+c_{ij}(1-p)$ if $w_{ij}$
exists too, or with probability $p(1-c_{ij})$ if it does not. In comparison to
the random networks considered above (Eq.\ \ref{wstats}), this connectivity
scheme preserves the mean weight $\bar{w}\equiv \left\langle w_{ij} \right\rangle =
\pm p w_0 / \sqrt{N}$ as well as the weight variance
$\left\langle (w_{ij}-\bar{w})^2 \right\rangle = p(1-p)w_0^2/N$ while giving
full control over their normalized covariance $c_{ij}$.
Note that $c_{ij}$ can assume positive values as high as $c_\text{max}=1$, in
which case all connections are bidirectional. However, $c$ cannot go below
$c_\text{min} = -p/(1-p)$, which stems from the sparsity condition that imposes
a certain degree of symmetry in $\tmmathbf{W}$: because both $w_{ij}$ and
$w_{ji}$ are zero with high probability, they will often be null together,
meaning that they cannot be fully anti-correlated. The limit case $c =
c_\text{min}$ corresponds to the complete absence of reciprocal connections.
Since we aim at tilting $\tmmathbf{W}$ towards antisymmetry, we choose $c_{ij}
= \kappa c_\text{min}$ when neurons $i$ and $j$ are of the same type, and
$c_{ij} = \kappa c_\text{max}$ when the two neurons have different types.  Thus
$0<\kappa<1$ parameterises the degree of antisymmetry in $\tmmathbf{W}$. As can
be seen in Fig.\ \ref{fig6}a, increasing $\kappa$ effectively decreases the
spectral abscissa $R^\prime = \max_\lambda \text{Re}(\lambda)$, although it is
designed not to affect the overall connectivity ``power'' $\sum w_{ij}^2$ which
\emph{is} the relevant quantity for amplification. Thus, for a fixed level of
dynamical slowing (i.e.  fixed $R^\prime$), antisymmetric connectivity matrices
can assume larger weight strengths and thereby yield stronger nonnormal
amplification than their random counterparts, as depicted in Fig.\
\ref{fig6}\textbf{b}.  Finally, note that a matrix with $\kappa=1$ is
\emph{not} purely antisymmetric ($\tmmathbf{W}^\dag \neq -\tmmathbf{W}$).  In
fact, neural connectivity matrices can never be fully antisymmetric, because
of the constraint that neurons can be only excitatory or only inhibitory. This
is an advantageous restriction here, because a fully antisymmetric matrix --
just like a fully symmetric one -- is in fact a normal operator that cannot
support transient amplification.

\section{Discussion}

The nonnormal nature of the neuronal connectivity could play a major role in
the functional dynamics of cortical networks. It can allow fast transients to
develop along well-defined activity motifs stored in the pattern of synaptic
efficacies. In networks with locally dense connectivity, the balance between
excitation and inhibition has been shown to generate such type of
amplification, accordingly termed ``balanced amplification'' \cite{Murphy09}.
We have revisited this feature in sparse balanced networks in which any two
neurons are connected randomly with some probability. Random networks had
already been studied in terms of their pseudospectrum \cite{Trefethen05}, which
only provides bounds on amplification. We have chosen a more direct approach
and assessed nonnormality in terms of its functional impact in networks driven
by stochastic external input. We have explicitly calculated the strength of the
activity fluctuations that can only be attributed to the nonnormality of the
recurrent connectivity.  We found nonnormal amplification to be very weak,
concluding that the only way to obtain large amplification in random networks
is to allow for significant dynamical slowing. If the dynamics are to be kept
fast, then the connectivity needs some structuring, so as to allow synaptic
weights to take up larger values and to discourage the emergence of large
positive eigenvalues.  We have given an example of minimal network structure,
namely connection antisymmetry, that achieves precisely this.  More adaptive
ways of shaping the connectivity, such as synaptic plasticity, could also be
considered. In particular, inhibitory synaptic plasticity has recently been
shown to suppress the attractor dynamics of a few activity motifs embedded in a
spiking network, while still permitting their transient recall \cite{Vogels11}.

Nonnormal amplification could provide a mechanistic account for the often
reported transient nature of both spontaneous and evoked activity in primary
sensory cortices.  Moreover, from a functional viewpoint, amplification without
slowing could be a highly relevant feature in areas involved in the processing
of fast-changing signals.  If transient amplification by the synaptic
connectivity is meant to allow past experience to be reflected in the responses
to sensory stimuli (see e.g. \cite{Fiser10}), then it is quite reassuring that
random networks are poor amplifiers, for it implies that nothing can be
amplified that has not been learned.
 
Here we have focused on spontaneous activity, i.e. on the fluctuations elicited
by isotropic external noise that is totally uninformed of the frozen structure
of the connectivity matrix. The equivalent triangular form of a nonnormal
connectivity matrix suggests that neuronal networks should be more sensitive
along some input directions than along others, so they could still respond
vigorously though transiently to some carefully chosen input patterns (evoked
activity).  The first Schur mode, for example, is indeed such a preferred
pattern \cite{Ganguli08}.  This anisotropy prompts two important questions.
First, how many different (orthogonal?) directions of high sensitivity does a
network possess?  Similarly, in how many distinguishable directions can the
network amplify those preferred input signals? These quantities taken together
could define the ``nonnormal information capacity'' of a network, reminiscent
of the concept of memory capacity in attractor networks.
 
Finally, our analysis has revealed that the nonnormality of balanced networks
is to a large extent reflected in large ``DC'' fluctuations.  This seems to be
a general feature of networks in which neurons can either be excitatory or
inhibitory, but not of a mixed type \cite{Kriener08}.  It is somewhat
disappointing that however strong activity fluctuations are in individual
neurons, they always comprise a finite fraction of common variability. This is
because the variance of the overall population activity is of the same order as
the activity variance of the individual neurons (Eq.\ \ref{dc}).  Should
computations exploit the fluctuations along the remaining $N-1$ degrees of
freedom of the network, complications in decoding the current network state
would most certainly arise from a single dimension dominating the dynamics.
However, we wish to point out that these large DC fluctuations are in fact a
direct consequence of the \emph{exact} excitation-inhibition balance considered
here. We show in Appendix \ref{eibalance} that when inhibition \emph{dominates}
over excitation, the variance of the population activity becomes suddenly
inversely proportional to the network size.  Furthermore, the mean pairwise
correlation coefficient in the network scales similarly, and thus vanishes in
large networks unless the E-I balance is exact. Note that this phenomenon is
\emph{not} mediated by a destruction of the strong feedforward link from the
global balance disruption $\tmmathbf{d}$ onto the DC mode $\tmmathbf{v}$, as
described at the end of section \ref{link_schur_to_w}. Increasing the overall
amount of inhibition does preserve this strong link, but cancels its amplifying
effect by imposing an equally strong negative feedback from the DC mode onto
itself (see Appendix \ref{eibalance}).  This dynamic cancellation of
fluctuations and correlations was already shown to arise in balanced networks
of spiking neurons \cite{Renart10}. Our results obtained for linear networks
therefore suggest it may be a very general feature of inhibition-dominated
balanced networks, and that fine-tuning the balance until it becomes exact
\cite{Vogels11} may strongly affect the dynamics of the network and the
resulting correlation structure.

\subsection*{Acknowledgments}
 
G.H. was partially supported by the European Union
Framework 7 ICT Project no.  215910 (BIOTACT, \texttt{www.biotact.org}). T.P.V.
was supported by the European Community's Seventh Framework Marie Curie
International Reintegration grant no. 268436.  G.H and T.P.V. were also
partially supported by BrainScaleS, an EU FET-Proactive FP7 project under grant
no.  269921.  Thanks to L. Abbott, J. de la Rocha, K. Miller and Y. Ahmadian
for helpful discussions.

\appendix

\section{Amplification in random triangular networks}
\label{appendix}

In this appendix we derive an exact expression for amplification in random
strictly triangular networks with linear stochastic dynamics as in Eq.\
\ref{ouprocess}, where the non-zero elements of the coupling matrix
$\tmmathbf{T}$ are drawn from an arbitrary distribution with zero mean and
variance $\alpha^2/N$ where $N$ is the network size.
Though no closed-form solution is known for the
zero time lag covariance matrix $\tmmathbf{\Sigma}$, we know from the theory of
multidimensional Ornstein-Uhlenbeck processes that it satisfies the so-called
Lyapunov equation \cite{Gardiner85}
\begin{equation}
  \left( \tmmathbf{T} - \mathbb{1} \right) \tmmathbf{\Sigma} +
  \tmmathbf{\Sigma} \left( \tmmathbf{T}^{\dag} - \mathbb{1} \right) = - \tau
  \sigma_{\xi}^2 \mathbb{1} \label{sigma}
\end{equation}
Equating component $(i,j<i)$ on both sides of Eq.\ \ref{sigma} yields:
\begin{equation}
  \sigma_{i j} = \frac{1}{2} \sum_{k = 1}^{i-1} t_{ik}
 \sigma_{j k} + \frac{1}{2} \sum_{k = 1}^{j - 1} t_{jk} \sigma_{ik} \label{firstequation}
\end{equation}
and equating the diagonal term $(i,i)$ on both sides gives the variance of Schur
mode $i$:
\begin{equation}
  \sigma_{ii} = \frac{\tau\sigma_\xi^2}{2}
 + \sum_{j = 1}^{i-1} t_{ij} \sigma_{ij} \label{secondequation}
\end{equation}
Combining Eqs.\ \ref{firstequation} and \ref{secondequation} yields
\begin{equation}
  \sigma_{ii} = \frac{\tau\sigma_\xi^2}{2}
  + \frac{1}{2} \sum_{j = 1}^{i-1} t_{ij}
  \left( \sum_{k = 1}^{i-1} t_{ik} \sigma_{j k} + \sum_{k =
  1}^{j - 1} t_{j k} \sigma_{ik}  \right)
  \label{combination}
\end{equation}
in which $\sigma_{j k}$ and $\sigma_{ik}$ are to be
recursively obtained from Eq.\ \ref{firstequation} with proper
replacement of indices. We would like
to calculate the expected value over the $t_{i j}$ coefficients, i.e. over
multiple realisations of random matrix $\tmmathbf{T}$. Explicitly
expanding the sums will reveal cross-terms like $\left\langle t_{i j}
t_{k \ell} \right\rangle$. Those vanish
if $i \neq k$ or $j \neq \ell$, because the coupling coefficients
are taken to be uncorrelated. The only remaining terms will be powers of
the variance $\alpha^2 / N$. Here we
seek a truncation to order $\alpha^4 / N^2$. Let us calculate:
\begin{equation}\label{firststep}
  \begin{split}
  \left\langle \sigma_{ii}
  \right\rangle = \frac{\tau\sigma_\xi^2}{2} + \frac{1}{2} \sum_{j = 1}^{i-1}
  \sum_{k = 1}^{i-1} \left\langle t_{ij} t_{ik} \sigma_{j k} 
\right\rangle  \\
+  \frac{1}{2} \sum_{j = 1}^{i-1} \sum_{k =
  1}^{j - 1} \left\langle t_{ij} t_{jk} \sigma_{ik} \right\rangle
\end{split}
\end{equation}
Because the network of Schur modes is purely feedforward, the cross-covariance
$\sigma_{j k}$ for $\left( j, k \right) < i$ is independent of the
coupling coefficients $t_{ij}$ and $t_{ik}$, thus
$\left\langle t_{ij} t_{ik} \sigma_{jk} \right\rangle = 
\left\langle t_{ij} t_{ik} \right\rangle
\left\langle \sigma_{jk} \right\rangle$.
The only non-vanishing term in the first double-sum is therefore
obtained for $k = j$, giving
\begin{equation}
  \left\langle \sigma_{ii}
  \right\rangle = \frac{\tau\sigma_\xi^2}{2} + \frac{\alpha^2}{2 N}
  \sum_{j = 1}^{i-1} \left\langle \sigma_{j j} \right\rangle + \frac{1}{2} \sum_{j
  = 1}^{i-1} \sum_{k = 1}^{j - 1} \left\langle t_{ij} t_{jk}
  \sigma_{ik} \right\rangle \label{firstaverage}
\end{equation}
Let us expand the expression in the second double-sum using Eq.\ \ref{firstequation}:
\begin{equation}\label{negligible}
  \begin{split}
  \left\langle t_{ij} t_{jk} \sigma_{ik}
 \right\rangle = \frac{1}{2} \sum_{\ell = 1}^{i-1} \left\langle t_{ij} t_{j k}
  t_{i\ell} \sigma_{k \ell}
  \right\rangle \\
 + \frac{1}{2} \sum_{\ell = 1}^{k - 1} \left\langle t_{ij} t_{j k}
  t_{k \ell} \sigma_{i\ell}
  \right\rangle
\end{split} 
\end{equation}
As above, the first sum vanishes except for $\ell = j$.
Should one continue and expand the second sum, one would receive
terms of order $\alpha^6 / N^3$ and more which are discarded here (see above).
Hence
\begin{equation}
  \left\langle t_{ij} t_{j k}  \sigma_{ik}
   \right\rangle = \frac{\alpha^2}{2 N} \left\langle t_{j k} \sigma_{j k}
  \right\rangle + \cdots \label{negligible2}
\end{equation}
Using similar arguments, we expand $\left\langle t_{j k} \sigma_{jk} \right\rangle$ to
order $\alpha^2/N$ and receive:
\begin{equation}
  \left\langle t_{j k} \sigma_{j k} \right\rangle = \frac{\alpha^2}{2 N}
  \left\langle \sigma_{k k} \right\rangle + \cdots
\end{equation}
From Eq.\ \ref{firstaverage} it therefore follows that
\begin{equation}
\begin{split}
  \left\langle \sigma_{ii}
  \right\rangle = \frac{\tau\sigma_\xi^2}{2} &+ \frac{\alpha^2}{2 N}
  \sum_{j = 1}^{i-1} \left\langle \sigma_{j j} \right\rangle \\
 &+ \frac{\alpha^4}{8
  N^2} \sum_{j = 1}^{i-1} \sum_{k = 1}^{j - 1} \left\langle \sigma_{k k}
  \right\rangle
\end{split}
\end{equation}
Defining $f_i= 2\left\langle \sigma_{i i} \right\rangle /(\sigma_\xi^2 \tau)$,
we end up with a recursive equation for the
build-up of relative variance down the feedforward network of Schur modes:
\begin{equation}\label{buildingup}
  f_i = 1 + \frac{\alpha^2}{2 N} \sum_{j = 1}^{i -
  1} f_j + \frac{\alpha^4}{8 N^2} \sum_{j = 1}^{i - 1} \sum_{k = 1}^{j - 1}
  f_k 
\end{equation}
Now we define $x=i/N$ (thus $0 \leq x \leq 1$) and rewrite Eq.\
\ref{buildingup} as
\begin{equation}
  \begin{split}
  f_{xN} = 1 &+ 
  \frac{\alpha^2 x}{2 i} \sum_{j = 1}^{i - 1} g \left( 
  \frac{xj}{i} \right) \\
  & + \frac{\alpha^4 x^2}{8 i^2} \sum_{j = 1}^{i - 1}
  \sum_{k = 1}^{j - 1} g \left(\frac{xk}{i} \right)
\end{split}
\end{equation}
In the limit $N\to\infty$ with constant $x=i/N$ ratio, 
the sums on the r.h.s.
converge to their corresponding Riemann integrals,
endowing $f_{xN}$ with a proper limit $g(x)$:
\begin{equation}\label{integralequation}
\begin{split}
  g \left( x \right) = 1 &+ \frac{\alpha^2 x}{2} \int_0^1 g \left(
  x s \right) d s \\
& + \frac{\alpha^4 x^2}{8} \int_0^1 d s \int_0^1 d
  s' \Theta \left( s - s' \right) g \left( x s' \right) 
\end{split}
\end{equation}
where $\Theta$ is the Heaviside function. This convergence stems from the $1/N$
scaling of the variance $\alpha^2/N$.  Using straightforward changes of
variables ($s \mapsto s/x$), we end up with an integral equation for $g$,
the continuous variance profile along the (now infinitely large) network of
Schur patterns: \begin{equation}
  g \left( x \right) = 1 + \frac{\alpha^2}{2} \int_0^{x} g \left(
  s \right) d s + \frac{\alpha^4}{8} \int_0^{x} d s \int_0^s d s' g
  \left( s' \right) \label{finalintegralequation}
\end{equation}
Differentiating Eq.\ \ref{finalintegralequation} twice with respect to
$x$ yields a second-order differential equation for $g$
\begin{equation}
  g'' \left( x \right) = \frac{\alpha^2}{2} g' \left( x \right) +
  \frac{\alpha^4}{8} g \left( x \right) 
\end{equation}
with initial conditions $g(0)=1$, $g'(0)=\alpha^2/2$,
and $g''(0)=3\alpha^4/8$. The solution is precisely $g^{\text{\tiny{LB}}}(x)$ given in Eq.\
\ref{secondorderprofile} of the main text. It is only a lower-bound on the
true variance profile $g(x)$ since all the higher-order terms in $\alpha^2$
that we have neglected are positive.
This approximation proves reasonable for $\alpha^2<3$ as shown
in Fig.\ \ref{fig3}\textbf{a} (dashed blue lines).
Further integrating over $x$ yields a lower-bound
on nonnormal amplification $A_0(\alpha^2) \equiv \int_0^1 g \left( x \right) dx -1$
(Fig.\ \ref{fig3}\textbf{b}, dashed blue line):
\begin{equation}\label{secondorder}
\begin{split}
  A_0^{\text{\tiny{LB}}} \left( \alpha^2 \right) &= 
  \frac{2}{\alpha^2 \sqrt3} \exp \left( -\frac{(\sqrt3-1)\alpha^2}{4} \right) \\
  &\times  \left[
\exp\left(\frac{\sqrt3 \alpha^2}{2} \right) - 1
\right] - 1
\end{split}
\end{equation}
Instead of truncating $\left\langle\sigma_{ii}\right\rangle$ to order
$\alpha^4$, one can also decide to start again from Eq.\ \ref{firststep} and
keep all terms up to order $n$. This requires careful counting, and results in
a differential equation of order $n$, reading
\begin{equation}\label{ordern}
g^{(n)}(x) = \frac{\alpha^2}2\sum_{k=0}^n C_k\left(\frac{\alpha^2}4\right)^k 
g^{(n-k-1)}(x)
\end{equation}
where $C_k=(2k)!/\left[k!(k+1)!\right]$ is the $k^{\text{th}}$ Catalan number.
Assuming $g(x)$ can be written for $0\leq x \leq 1$ as a convergent power series
\begin{equation}
\label{powerseries_appendix}
g(x)=\lim_{K\to\infty}\sum_{k=0}^K \beta_k x^k
\end{equation}
and equating $g^{(k)}(0)$ in both Eqs.\ \ref{ordern} and \ref{powerseries_appendix} yields
the results of Eqs.\ \ref{powerseries} -- \ref{finalamp}.

\bigskip

\section{Variance of the DC component}
\label{appendixuniformmode}
                                      
The last Schur mode is fed by the activities of all previous Schur vectors,
weighted by couplings with variance $\zeta_0^2/N$. The same calculation that led
to Eq.\ \ref{buildingup} in this case leads to
\begin{equation}\label{variancefirstschurmode}
  f_N = 1 + \frac{\zeta_0^2}{2} \sum_{j = 1}^{N-1}
 f_j + \frac{\zeta_0^2 R^2}{8N} \sum_{j = 1}^{N-1} \sum_{k = 1}^{j-1}
  f_k + \cdots
\end{equation} 
which can be rewritten as
\begin{equation}
\frac{f_N}{N} = \frac{1}{N} + \frac{\zeta_0^2}{R^2} \left(
   \frac{R^2}{2N} \sum_{j = 1}^{N-1} f_j + \frac{R^4}{8N^2}
  \sum_{j=1}^{N-1} \sum_{k=1}^{j-1} f_k +\cdots\right)
\end{equation}
where the sums were previously calculated in the limit $N\to\infty$
(Eqs.\ \ref{buildingup} -- \ref{powerseries_appendix}).
We thus recover
\begin{equation}
\label{amplastschurmode}
\lim_{N \rightarrow \infty} \frac{f_N}{N} = \frac{\zeta_0^2}{R^2} 
   \left[ g \left( 1 \right) - 1 \right] 
\end{equation}
With $\zeta_0^2$ given by Eq.\ \ref{varfirstline} we arrive at
Eq.\ \ref{dc} of the main text.

\section{\label{eibalance}Exactly balanced vs. inhibition-dominated networks}

In this paper, we have considered connectivities in which weights were either
zero or $\pm w_0/\sqrt{N}$, the $\pm$ sign depending on the excitatory vs.
inhibitory nature of the presynaptic neuron (Eq.\ \ref{wstats}). Furthermore,
the number of cells of both types was identical. The total inhibitory synaptic
strength thus exactly matched its excitatory counterpart.
In this appendix, we wish to show that if the non-zero inhibitory weights are stronger,
i.e. $-\gamma w_0/\sqrt{N}$ with $\gamma>1$, the dynamics of the overall
population activity is strongly affected.

\begin{figure}[tb!]
\bigskip
\centering
\includegraphics[width=3.4in]{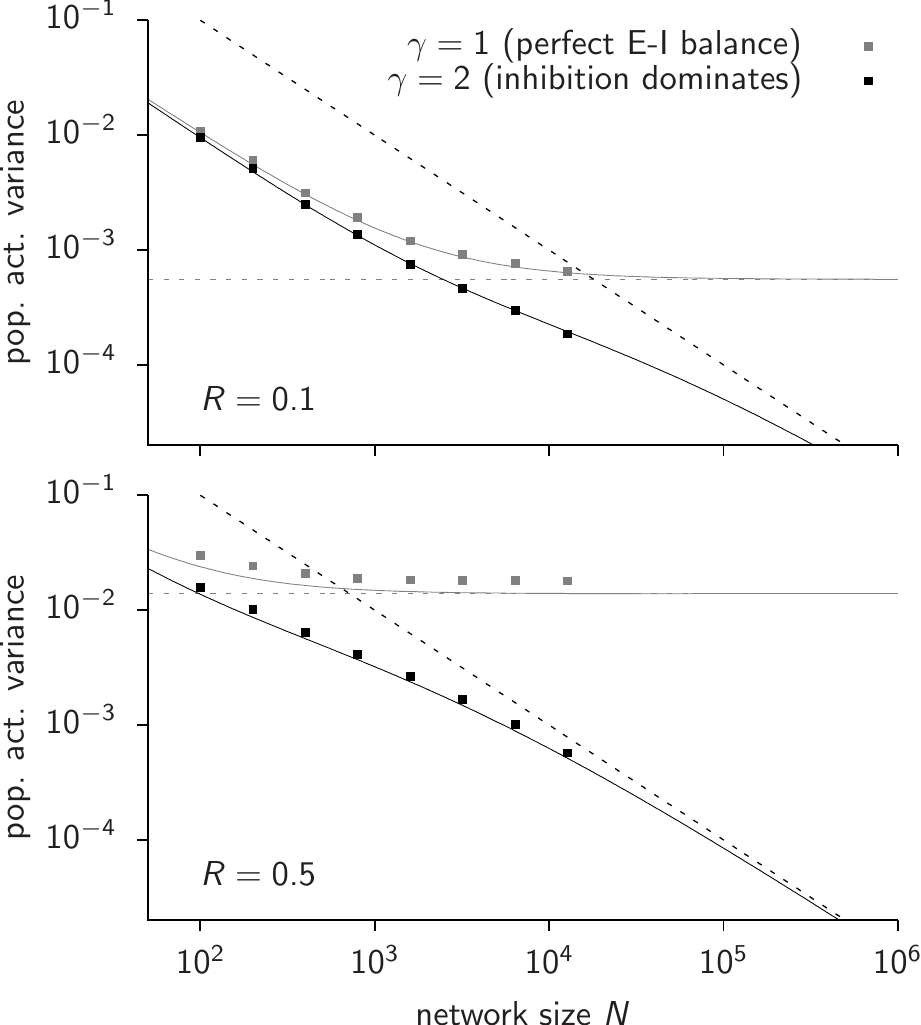}
\caption{\label{fig7}Temporal fluctuations of the overall population firing
rate in a balanced neuronal network.  The variance of the average population
activity $\mu(t)=\sum x_i(t) /N$ is reported as a function of the network size $N$,
in logarithmic scale. When inhibition perfectly balances excitation
($\gamma=1$), the variance is asymptotically independent of the network size
(gray). When inhibition dominates ($\gamma>1$), it scales with $1/N$ (black).
The solid lines denote the approximation in Eq.\ \ref{vardcgamma}. The dashed
lines indicate the asymptotics (Eq.\ \ref{asymptotics}).
Points denote the empirical variance obtained by simulating Eq.\ \ref{ouprocess}
for 100 seconds, for a neuronal network constructed as specified in section \ref{link_schur_to_w}
with connectivity density $p=0.1$. The spectral radius was set to
$R=0.1$ (top plot) and $R=0.5$ (bottom plot).
}
\end{figure}

We have seen that the ``DC'' mode $\tmmathbf{v}=(1,\ldots,1)/\sqrt{N}$ is an eigenvector
of $\tmmathbf{W}$. Let $\lambda_v$ denote the associated eigenvalue,
which quantifies the effective decay rate of the DC component in the network
of neurons. If the E-I balance is exact ($\gamma=1$) as assumed throughout the paper,
then $\lambda_v=0$. More generally, however, one can calculate
\begin{equation}\label{lambdav}
\lambda_v \quad = \quad - \frac{p w_0(\gamma-1)}2 \cdot \sqrt{N}
\end{equation}  
We see there is an unexpected scaling that the exact balance was hiding :
$-\lambda_v \sim \mathcal{O}(\sqrt{N})$.
Note that all other eigenvalues are now scattered inside the disk of radius
\begin{equation}\label{newr}
R\quad=\quad w_0 \sqrt{\frac{(1+\gamma^2) p(1-p)}2}
\end{equation}
though no longer uniformly so since the variance of the inhibitory and excitatory weights
now differ by a factor of $\gamma^2$ \cite{Rajan06}.
Having kept the focus of this paper on nonnormal effects, we have intentionally
set aside the contributions of the eigenvalues to the overall amplification in
the network. When $\lambda_v = 0$ (perfect balance), our prediction that the
average population activity $\mu(t) \equiv \sum x_i(t) /N$ should have a
variance of order $\mathcal{O}(1)$ was justified : the last Schur unit corresponding
to this DC indeed receives $N-1$ contributions of order $\mathcal{O}(1)$, and
its decay time constant is simply $\tau \sim \mathcal{O}(1)$, yielding
$\textrm{var}[\mu(t)] \sim \mathcal{O}(1)$.  When inhibition dominates
($\gamma>1$), the DC component suppresses itself via a negative feedback
that scales with $\sqrt{N}$, yielding a very short decay time constant
$\tau/(1-\lambda_v) \sim \mathcal{O}(1/\sqrt{N})$ whose deviation from $\tau$
can no longer be neglected.  To see what the implications of this scaling are
for the variance of $\mu(t)$, let us reduce the dynamics of the DC to the
following set of $N$ stochastic differential equations:
\begin{equation}\label{reduceddcdynamics}
\begin{array}{rcl}
dy_i &=& \displaystyle -\frac{dt}\tau y_i + \sqrt\frac{2}{\tau} \, d\xi_i \quad \textrm{for } \quad 1\leq i<N\\
dy_N &=& \displaystyle \frac{dt}\tau\left(-(1-\lambda_v) y_N + \sum_{i=1}^{N-1} \varepsilon_i x_i\right)
+\sqrt\frac{2}{\tau} d\xi_N
\end{array}
\end{equation}
Here $y_1,\ldots,y_{N-1}$ model the first $N-1$ Schur units independently, with the appropriate
noise terms such that they achieve a variance of one (corresponding to the limit of small
amplification). They feed $y_N$ -- which models the activity of the last Schur unit, i.e. the
DC component $\mu(t)\sqrt{N}$ -- with couplings $\varepsilon_i$ such that $\sum \varepsilon_i^2/N = \zeta_0^2$.
We calculate the coupling variance $\zeta_0^2$ the same way we did in section \ref{link_schur_to_w}:
\begin{equation}\label{newzeta0}
\zeta_0^2 \quad = \quad \frac{p^2 w_0^2 (1+\gamma^2)}2
\end{equation}
The variance $\textrm{var}[\mu(t)]$ of the overall neuronal population activity,
here modeled by $\mu(t)\approx y_N(t)/\sqrt{N}$, is given by standard Ornstein-Uhlenbeck theory:
\begin{equation}\label{vardcgamma}
\textrm{var}(\mu(t)) \quad = \quad \frac1{N(1-\lambda_v)}\left[
1+\frac{N\zeta_0^2}{2-\lambda_v}\right]
\end{equation} 
Although we have neglected amplification and correlations in the first $N-1$ Schur units,
Eq.\ \ref{vardcgamma} does provides a good intuition for how the mean population
activity $\mu(t) = y_N(t)/\sqrt{N}$ scales with the network size $N$, and provides
a good qualitative match to numerical results even for a non-negligible spectral radius $R=0.5$
(Fig.\ \ref{fig7}).

The asymptotics of $\textrm{var}[\mu(t)]$ are given by
\begin{equation}\label{asymptotics}
\textrm{var}[\mu(t)] \quad \sim \quad
\left\{ 
\begin{array}{ccl}
\displaystyle \frac{p^2 w_0^2}2 & \textrm{ if } & \gamma=1 \\
\displaystyle \frac{2(1+\gamma^2)}{N(\gamma-1)^2} & \textrm{ if } & \gamma>1
\end{array}
\right.
\end{equation} 

Thus, when inhibition dominates over excitation ($\gamma>1$), the fluctuations
of the overall population activity vanish for large networks, which was already
shown in \cite{Renart10} for inhibition-dominated networks of spiking neurons.
In contrast, fine tuning the connectivity such that the balance becomes exact
($\gamma=1$) opens the possibility for these fluctuations to subsist in
arbitrarily large networks.  This has profound consequences for the mean
pairwise correlation $\bar{r} \equiv \sum_{i\neq j}
\textrm{cov}[x_i(t),x_j(t)]/N^2$, as seen from the following identity
\begin{equation}\label{correlations}
\bar{r} \quad = \quad
\textrm{var}[\mu(t)] - \frac1{N^2} \sum_i \textrm{var}[x_i(t)]
\end{equation} 
We have seen that the average variance $\textrm{var}[x_i(t)]$ in the individual
neurons (i.e. amplification as we define it) is $\mathcal{O}(1)$. Thus, Eq.\
\ref{correlations} implies that $\bar{r}$
scales with $N$ in the same way $\textrm{var}[\mu(t)]$ does: either $\mathcal{O}(1)$
if the balance is perfect, or $\mathcal{O}(1/N)$ if inhibition dominates.

\bibliographystyle{unsrt}
\bibliography{mybib.bib}

\end{document}